\documentclass[a4paper,10pt]{article}
\usepackage[utf8x]{inputenc}
\title{A possible origin of the q=4/3 diquark}
\author{A. Rivero \thanks{Institute for Biocomputation and
Physics of Complex Systems, University of Zaragoza} \thanks{\texttt{al.rivero@gmail.com}, \texttt{arivero@unizar.es}}}

\begin{document}

\maketitle

\begin{abstract}
Between the new physics candidates proposed to explain the $t \bar t$ asymmetry measured 
in the Tevatron, there are 
some scalar diquarks with electric charge +4/3. This kind of diquark is also
needed to classify all the scalars  of 
the supersymmetric standard model, with three generations, under a global flavour symmetry.
\end{abstract}

\section{Introduction}
One of the measurements that have caused some stir during the last
run of the Tevatron 
has been the forward-backward asymmetry\cite{Aaltonen:2011kc} of top quark processes.
Between the different New Physics candidates that have been 
proposed (see \cite{AguilarSaavedra:2011ug} for a short review),
some scalar ``diquarks'' in diverse multiplets are favoured, and 
particularly \cite{arXiv:0912.0972} some  isosinglets with charge 4/3. 

While the hint is generically towards flavour models, here we want to show that
when flavour is imposed not in the standard model fermions
but in the scalar sector (squarks and sleptons) of the supersymmetric SM, the 
candidate particles appear 
inside a very unique construction, that exhausts exactly for three generations with
five light quarks. We explain this fact in the first section of the paper and then
we proceed to speculate on other uses of a flavour symmetry inspired in composites.

\section{Flavour in susy scalars}

Independently of its physical interpretation, it is possible to use a 
flavour $SU(5)$ to classify the 24 sleptons contained in the SSM -when extended with
right neutrinos-. This is done by taking this representation from $5 \otimes \bar 5 = 24 \oplus 1$
and then branching it down to $SU(3) \times SU(2)$.

\begin{equation}
 24=(1,1)+(3,1)+(2,3)+(2, \bar 3)+(1,8)
\end{equation}

Giving an electric charge $+2/3$ to the $SU(3)$ piece and $-1/3$ to $SU(2)$, it can be seen that
we have got a sextet of charge $+1$, another opposite sextet of charge $-1$, and another twelve states
of zero charge. The construction can be done for any odd number of generations, but it becomes
specially elegant -and unique in some terms- when done for three.

Now, we produce all the squarks from the same flavour group by taking the
$15$ of $\bar 5 \otimes \bar 5 = 15 \oplus 10$. With the same branching, it decomposes as
\begin{equation}
15=(3,1)+(2,3)+(1,6)
\end{equation}
so that we get one sextet of charge -1/3 and another sextet of charge +2/3. We call this
reproduction of the original charges a ``supersymmetric Bootstrap'', or {\em sBootstrap}
for short, because it is possible to interpret the $SU(3)$ piece as coming from 
three particles similar to quarks $d,s,b$ and the $SU(2)$ piece similar to quarks $u,c$. If we
pursue this interpretation, the uniqueness is more appealing: with more of three generations,
not all of them are used in the flavour symmetry, and the number of extra particles
makes the scheme a lot uglier, adding more exotic squarks and sleptons to the bag. Note
that also the same interpretation, with the same ``quarks'', applies to the 24 decomposition
on sleptons above.

This symmetry was found some years ago in \cite{Rivero:2005if} and the uniqueness is discussed with
more detail there, but even with three generations it had a 
severe handicap: the $15$ multiplet
has three extra scalars not in the standard model. Fortunately, they were different from
the other scalars in the sense that, being a odd number, it was not possible to arrange them in
Dirac supermultiplets; then this chirality was expected to be an advantage
allowing either to eliminate them from the game board or to force them into the gauge sector (more
on this later). But an acceptable, fully developed solution has not been found yet. 

What is important for this brief letter is to notice what this $(3,1)$ triplet is: the three 
scalars have charge +4/3 and can be assigned interactions as a scalar diquark. It is then
the same kind of particle expected to solve the Tevatron asymmetry.

We can read this in two ways: on one direction, it could be said that an extra flavour symmetry
for the scalar sector of SUSY predicts the need of 4/3 diquarks. On the reverse direction, it
is possible that most of the models proposed to explain the asymmetry will exhibit explicitly
a scalar flavour symmetry when super-symmetrized. On the side of model builders, it may 
mean some extra work, because besides a colour triplet
we also have a flavour triplet (naively, $uu$, $uc+cu$ and $cc$ like states). The degeneracy
could be removed in some flavour-colour locking scheme, or it could be really there.

The initial idea in \cite{Rivero:2005if} was to interpret all those scalars as the
well known spectrum of mesons, plus a diquark spectrum hidden inside baryons. That implied 
some complications: besides being composites, the supersymmetry needed to work between particles
with different baryon and lepton number. Probably this is still the case here, but with
a revenge: if the +4/3 particles are real states on their own, and no just QCD diquarks, the
argument to keep insisting in the original idea becomes even weaker\footnote{I thank M. Porter
by this observation.}; the next resort is to consider all the scalar states as
formed by condensation via some technicolor force, beyond QCD but very similar to it.

\section{A mass spectrum.}

Having stated the main point, lets allow for an {\em Intermezzo}: If SUSY 
has a flavour symmetry of its own, based on composites, is a mass
formula possible?

It could be, and it could be that remnants of the mass formulae 
are still visible in the fermion sector even after SUSY breaking. This
idea was part of the motivation for this research, back in 2005, when
some attention was dedicated to Koide equation \cite{koide, koide2, Koide:1983qe}
and its repercussions in neutrino spectrum (e.g. \cite{Brannen:2010zza}. 
This equation is an evolution of the formulae for masses and Cabibbo angle that 
were justified in the late seventies using textures \cite{Fritzsch:2002fg},
democratic matrices, or permutation symmetry in $SU(2)_R \otimes SU(2)_L$ models\cite{Harari:1978yi}.

Recent empirical work \cite{Rodejohann:2011jj,arXiv:1111.7232} has extended 
Koide formula from leptons to quarks then bringing the equation back
to its birthplace. The $m_e=m_u=0$ spectrum, see \cite{arXiv:1111.7232} could be
an interesting starting point for the search of a supersymmetrical spectrum. Remember
that the LR models for Cabibbo angle used to permute the $R$ sector in a way that
the bottom R quark was related to the up L quark; if we can not put the 
superpartners of a $m_u=0$ quark at zero -say, if we do not want the lowest energy
state to be degenerated- then we can still put them at the energy of the $b$ quark.

\section{Towards technicolor}

Finally, we wonder if, besides to explain the $t \bar t$ asymmetry, is it 
possible to exploit these
diquarks in some Higgs-like scenario. The peculiar charge of these particles seems an
impediment, but there is an interesting possibility: that some of this charge comes
from $B-L$ in a non-chiral way. This is a typical feature of GUT models
using a left-right symmetry.

From this perspective, the $(3,1)$ diquarks should be seen as having some
non-chiral interactions, providing colour and a piece of $Q_v= \frac 12 (B-L)=\frac 12 . \frac 23 = \frac 13$ of
 electric charge, plus a chiral interaction providing the extant $Q_{ch}=+1$. If
the condensation mechanism can get rid of the vector part of the interaction,
our diquarks seem more alike the scalars of the gauge supermultiplets: color
neutral and with integer electric charge. Looking B-L in a different foot
that the rest of the symmetries is interesting because it also imply that we
increase our disbelief in R-symmetry. You could have noticed that the scalars we
are producing have a B-L value different in one unit respect to the fermions they are
supposed to partner with.

Furthermore, color neutrality causes a degeneration, and the number of different
states is now only three particles and the corresponding antiparticles. Under SUSY,
the massless gauge supermultiplet needs to eat one chiral supermultiplet (with a fermion
and two scalars) to get mass; one of the scalars becomes the extra degree of freedom
of the W or Z, and the other surfaces as a Higgs scalar. Thus it is interesting that
under color neutrality, and barring the degeneration -or controlling it with some
flavour-colour locking, appropriate also to the discussion in the first section-, we
get the number of scalars needed to make three chiral supermultiplets. Some work 
is being done to try to introduce these ingredients in an EWSB  
mechanism, but it is still in very early stages.

\section{Acknowedgements.}

I must thank updated information about the status of the
top asymmetry research, as well clarifications in the conventions, to Manuel Perez Victoria,
Nejc Kosnic and Ilja Dor\v{s}ner. Encouragement about the divulgation of the sBootstrap idea has been 
provided, direct or indirectly, by Tommaso Dorigo, Mitchell Porter and the mentors
of {\tt PhysicsForums.com}, who allowed for almost single handed development of some
threads.

%
% Y Koide, C. Brannen for the Koide eq.
%

\end{document}